# The Generalized Model of Polypeptide Chain Describing the Helix-Coil Transition in Biopolymers

Evgeni Sh. Mamasakhlisov[1], Artem V. Badasyan[*1,2],

Artyom V. Tsarukyan[1], Arsen V. Grigoryan[1], Vladimir F. Morozov[1].

[1]*Department of Molecular Physics, Yerevan State University, A.Manougian Str.1, 375025, Yerevan, Armenia.*

[2]*ICTP, Condensed Matter Section, Strada Costiera 11, 340014, Trieste, Italy.*

Abstract

In this paper we summarize some results of our theoretical investigations of helix-coil transition both in single-strand (polypeptides) and two-strand (polynucleotides) macromolecules. The Hamiltonian of the Generalized Model of Polypeptide Chain (GMPC) is introduced to describe the system in which the conformations are correlated over some dimensional range $\Delta$ (it equals 3 for polypeptide, because one H-bond fixes three pairs of rotation, for double strand DNA it equals to one chain rigidity because of impossibility of loop formation on the scale less than $\Delta$). The Hamiltonian does not contain any parameter designed especially for helix-coil transition and uses pure molecular microscopic parameters (the energy of hydrogen bond formation, reduced partition function of repeated unit, the number of repeated units fixed by one hydrogen bond, the energies of interaction between the repeated units and the solvent molecules). To calculate averages we evaluate the partition function using transfer-matrix approach. The GMPC allowed to describe the influence of a number of factors, affecting the transition, basing on a unified microscopic approach. Thus we obtained, that solvents change transition temperature and interval in different ways, depending on type of solvent and on energy of solvent-macromolecule interaction; stacking on the background of H-bonding increases stability and decreases cooperativity of melting. For heterogeneous DNA we could analytically derive well known formulae for transition temperature and interval. In the framework of GMPC we calculate and show the difference of two order parameters of helix-coil transition - the helicity degree, and the average fraction of repeated units in helical conformation. Given article has the aim to review the results obtained during twenty years in the context of GMPC.

corresponding author's e-mail: abadas@www.physdep.r.am

Section 1. Introduction

Since the 1960's the helix - coil transition in biopolymers has been a topic of intensive investigations [1-29] and is still vigorously discussed [30-52]. Traditionally the theoretical models for the transition assume that each base pair can be in either the helical or the coil state. This assumption makes it convenient to use the Ising model [1-3, 5-7, 23-29] or to calculate the free energy directly as though the system were a dilute one-dimensional solution of helix and coil junctions [14, 15]. While different in details, most traditional theories use the mean-field approximation. In other words, the Hamiltonians of these models use parameters that are averages over all conformations of the molecule (e.g. the cooperativity parameter in Zimm - Bragg (Z-B) theory [23-26] or the junction free energy [14, 15]). Among few microscopic approaches to the helix-coil transition the theory of Lifson, Roig (L-R model) and others for polypeptides is widely known and used [1, 27-29]. The microscopic Generalized Model of Polypeptide Chain (GMPC), that we discuss and develop here, includes L-R model as particular case. It is well-known, that the simplification of L-R model, in its turn, results in Z-B model. Thus, the hierarchy of models is established (from the mostly achieved complete to the simplest one) GMPC→Lifson-Roig model→Zimm-Bragg model.

From the very beginning we want to say, that we are not neglecting or diminishing the huge role of these two beautiful and elegant models (Z-B and L-R) at all, but are instead developing these approaches and bringing the phenomenon of helix-coil transition into the realm of modern statistical physics to achieve the mostly up-to-date possible complete description of phenomenon.

The GMPC allowed to describe the influence of a number of factors, affecting the transition, basing on a unified microscopic approach. Thus, important heterogeneity of DNA and polypeptide (PP) structural composition, stacking interactions (on the background of hydrogen bonding), solvent influence and many others were taken into account. Many old, classical results are reevaluated and checked and some new ones are obtained [53-58] in scope of GMPC in past 20 years. Given article has the aim to review these results.

The structure of article is as follows. This small Introduction (Section 1) is in the beginning. The basic GMPC "vacuum" model is presented in short in Section 2. Then the models, accounting for different factors on the basis of GMPC are reviewed (Section 3). Section 4 discusses the definition of order parameter. Obtained results are concluded in Section 5.

Section 2. Basic model (GMPC) for helix-coil transition description.

A microscopic Potts-like one-dimensional model with $\Delta$-particle interactions describing the helix – coil transition in polypeptides was developed in [53-55]. Then it was shown that the same approach could be applied to DNA if ignore large-scale loop factor [56]. The Hamiltonian of GMPC has the form of the sum over all repeated units:

$$-\beta H = J\sum_{i=1}^{N}\delta_i^{(\Delta)}, \qquad (1)$$

where $\beta = T^{-1}$ is inverse temperature; $N$ is the number of repeated units; $J=U/T$ is the temperature-reduced energy of interchain hydrogen bonding; $\delta_j^{(\Delta)} = \prod_{k=\Delta-1}^{0}\delta(\gamma_{j-k},1)$, with Kronecker $\delta(x,1)$; $\gamma_l$ is spin that can take on values from 1 to $Q$ and describes the conformation of $l$-th repeated unit. The case when $\gamma_l$ is equal to 1 denotes the helical state, other ($Q$-1) cases correspond to coil state and thus the conformational freedom of coil is described. $Q$ is the number of conformations of each repeated unit. The Kronecker delta inside the Hamiltonian ensures that energy $J$ emerges only when all $\Delta$ neighboring repeated units are in helical conformation.

The transfer - matrix, corresponding to the Hamiltonian Eq.(1) looks like:

$$\hat{G}(\Delta) = \begin{pmatrix} V & V & V & ... & V & V & V \\ 1 & 0 & 0 & ... & 0 & 0 & 0 \\ ... & ... & ... & ... & ... & ... & ... \\ 0 & 0 & 0 & ... & 1 & 0 & 0 \\ 0 & 0 & 0 & ... & 0 & 1 & Q \end{pmatrix}, \qquad (2)$$

where all elements of first row are equal to $V=\exp[J]-1$; all elements of first lower pseudodiagonal are 1; the $(\Delta,\Delta)$ element is $Q$; all other elements are zero. The secular equation for this matrix looks like:

$$\lambda^{\Delta-1}[\lambda - (V+1)](\lambda - Q) = V(Q-1). \qquad (3)$$

As previously shown [55], the two-particle correlation function of this model in thermodynamic limit can be written as

$$g_2(r) = \left\langle \delta_i^{(\Delta)} \delta_{i+r}^{(\Delta)} \right\rangle - \left\langle \delta_i^{(\Delta)} \right\rangle \left\langle \delta_{i+r}^{(\Delta)} \right\rangle \sim \exp\left[-\frac{r}{\xi}\right], \qquad (4)$$

where $r$ is the distance (in repeated units), and

$$\xi = \left[\ln \lambda_1 / \lambda_2 \right]^{-1} \qquad (5)$$

is the correlation length; $\lambda_1$ is the largest eigenvalue, $\lambda_2$ is the second largest. Near the transition point, estimated from $W=Q$ condition as $T_m = U / \ln Q$, the correlation length $\xi$ passes through the maximum, which can be estimated as [53-55]

$$\xi_{\max} \sim Q^{\frac{\Delta-1}{2}}. \qquad (6)$$

The following set of parameters was evaluated for DNA in Ref.[56]: $Q \propto 3 \div 5$; $\Delta \propto 10 \div 15$. For polypeptides $Q \propto 60 \div 90$; $\Delta = 3$.

Now as to the relation of GMPC to Lifson-Roig and Zimm-Bragg theories. It is easy to see, that the secular equation Eq.(3) for $\Delta = 3$ exactly coincides with that of Lifson-Roig model. The parameters of Zimm-Bragg theory are linked to GMPC parameters [56] as

$$\sigma = \xi_{\max}^{-2}, \quad s = W/Q = \frac{\exp[U/T]}{Q}. \qquad (7)$$

One can see that for DNA set the cooperativity parameter $\sigma$ varies from $10^{-5}$ to $10^{-7}$, while for polypeptide set it is about $10^{-2}$. So, the high cooperativity of homogeneous DNA was explained as determined by large value of $\Delta$ which reflects the high rigidity of one-strand DNA.

To summarize we want to say that in our GMPC the following features of system are modeled:

1. conformational ability of each repeated unit (entropic advantage of coil state) through the parameter $Q$;
2. the scale of restrictions, imposed by one hydrogen bonding through the parameter $\Delta$;
3. the energy of one hydrogen bonding through the parameter $V = e^J - 1$.

Section 3. Some factors influencing the transition

Subsection 3.1. Solvent

As is known, biopolymers in cells exist in presence of water, in which ions of metals and other low and high molecular weight compounds are dissolved [1-3, 59]. Experiments on biopolymer melting are as well conducted in presence of solvents of different types, mainly containing water [16-22]. The solvents by their molecular weight can be divided onto two groups: high and low molecular, and by binding type: reversibly (ligands) and irreversibly binding. If classify reversibly binding solvents by the mechanism of binding with biopolymers, we obtain two main groups: competing for hydrogen bond formation and non-competing. We perform the account of solvent influence for low-molecular, reversibly binding solvents, both competing and non-competing.

*3.1.1. The model of competing solvent*

Molecules of some solvents, e.g., water are able to establish hydrogen bonds with nitrogen bases of DNA or peptide groups of polypeptides [1-3]. Thus we construct the Hamiltonian of competing solvent model basing on GMPC and taking into account, that [54, 55]:

1) If repeated unit is not bonded by intramolecular H-bond, it is free for intermolecular polymer-solvent H-bond formation.

2) H-bonds between polymer and solvent molecules are formed or not depending of the orientation (state) of solvent molecule with respect to repeated units of biopolymer. Orientation of solvent molecules are considered to be discrete and the number of orientations is $q$.

3) To each orientation of solvent molecule a spin variable $\mu_j$ is put into correspondence, attaining values from 1 to $q$. State number 1 corresponds to bound state with energy $E$.

4) In case of polypeptide one hydrogen bond is formed in each repeated unit and, therefore, there is one binding site for solvent molecule; in case of DNA there are two or three h-bonds in each repeated unit and in case these bonds are broken, solvent molecule has four or five binding sites on nitrogen bases. Therefore let us consider general case with $m$ binding sites.

The Hamiltonian of such system looks like

$$-\beta H = -\beta H_0 + \beta E \sum_{i=1}^{N}(1-\delta_i^{(\Delta)})\sum_{j=1}^{m}\delta(\mu_j,1), \qquad (8)$$

where $H_0$ is the Hamiltonian Eq.(1); $E$ is the energy of intermolecular H-bond between polymer and solvent; variables $\mu_j=1,2,...q$ describe the orientation of solvent molecules with respect to repeated units of biopolymer.

The partition function of such system is

$$Z = \sum_{\{\gamma_{ij}\}}\exp[-\beta H] = \sum_{\{\gamma_{ij}\}}\exp[J\sum_{i=1}^{N}\delta_i^{(\Delta)} + I\sum_{i=1}^{N}(1-\delta_i^{(\Delta)})\sum_{j=1}^{m}\delta(\mu_j,1)] =$$

$$= \sum_{\{\gamma_i\}}\prod_{i=1}^{N}\exp\left[J\delta_i^{(\Delta)}\right]\sum_{\{\mu_j\}}\prod_{j=1}^{m}\exp\left[I\left(1-\delta_i^{(\Delta)}\right)\delta(\mu_j,1)\right],$$

where $\{\gamma_{ij}\} = \{\gamma_{i,polymer}\}\cup\{\mu_{j,solvent}\}$, and $I = \beta E$, all other symbols has the same meanings as above.

If introduce designations $V = \exp[J] - 1; W = \exp[I] - 1$ the partition function is rewritten as

$$Z = \sum_{\{\gamma_i\}} \prod_{i=1}^{N} (1 + V\delta_i^{(\Delta)}) \sum_{\{\mu_j\}} \prod_{j=1}^{m} (1 + W(1 - \delta_i^{(\Delta)})\delta(\mu_j, 1)). \qquad (9)$$

Let us sum out the rhs expression by spin variables, characterizing solvent:

$$L_i \equiv \sum_{\{\mu_j\}} \prod_{j=1}^{m} \left[1 + W(1 - \delta_i^{(\Delta)})\delta(\mu_j, 1)\right] =$$

$$= \sum_{\mu_1=1}^{q} \sum_{\mu_2=1}^{q} \ldots \sum_{\mu_m=1}^{q} \left[1 + W(1 - \delta_i^{(\Delta)})\delta(\mu_1, 1)\right]\left[1 + W(1 - \delta_i^{(\Delta)})\delta(\mu_2, 1)\right] \times$$

$$\ldots \times \left[1 + W(1 - \delta_i^{(\Delta)})\delta(\mu_{m-1}, 1)\right]\left[1 + W(1 - \delta_i^{(\Delta)})\delta(\mu_m, 1)\right] =$$

$$= q^m + (1 - \delta_i^{(\Delta)})\left[mWq^{m-1} + C_m^2 W^2 q^{m-2} + \ldots + W^m\right] =$$

$$= q^m \delta_i^{(\Delta)} + (1 - \delta_i^{(\Delta)})[q + W]^m = [q + W]^m \left[1 - \delta_i^{(\Delta)} + \frac{q^m \delta_i^{(\Delta)}}{[q + W]^m}\right].$$

Thus we obtain, that the partition function of model with the competing solvent is

$$Z = (W + q)^{mN} \sum_{\{\gamma_i\}} \prod_{i=1}^{N} \left[1 + V\delta_i^{(\Delta)}\right]\left[1 - \delta_i^{(\Delta)} + \frac{q^m \delta_i^{(\Delta)}}{[q + W]^m}\right].$$

Let us rewrite the partition function as $Z = (W + q)^{mN} \cdot Z_0(\tilde{V})$, where $Z_0(\tilde{V})$ is the partition function of model without solvent, but with redefined parameter

$$\tilde{V} = \frac{q^m \exp[\frac{U}{T}]}{(\exp[\frac{E}{T}] + q - 1)^m} - 1. \qquad (10)$$

Multiplier $(W + q)^{mN}$ in Eq.(10) does not influence the calculation of thermodynamical characteristics of polymer and we can finally write

$$Z \propto \sum_{\{\gamma_i\}} \prod_{i=1}^{N} \left[1 + \tilde{V} \delta_i^{(\Delta)}\right], \tag{11}$$

Thus, all the averaged features of system with the Hamiltonian (8) as function of $\tilde{V}$ coincide with those of Hamiltonian as function of $V$. The temperature behavior of averaged parameters (helicity degree, for example) is determined by $\tilde{V}+1$ via $T$ dependence.

In Fig. 1 the temperature dependences of $\tilde{V}(T)+1$ at different positive values of $\alpha = \dfrac{mE}{U}$ are presented. The transition point is determined from the intersection of $\tilde{V}(T)+1$ and $Q$. Increasing the energy of intermolecular interaction shifts the transition point to the right. One can see that $\tilde{V}(T)+1$ dependence may loose its monotonic behavior, thus resulting in two transition points. One of them is helix-coil transition point. Another one corresponds to coil-helix transition at increasing temperature. We calculated the transition temperatures of both transitions and the results of these calculations are presented in Fig. 2.

The existence of upper bound for $\alpha$ for the $\alpha > 1$ case is conditioned by the convergence of iterations. This condition in explicit form looks like

$$\frac{1-1/q}{1-1/\alpha} + \frac{m}{\alpha} \ln \frac{(1/Q)^{1/m} - 1/q}{1-1/q} > 1.$$

Analyzing the curve, one can conclude, that the temperature of helix-coil transition decreases at increased energy of intermolecular H-bonding. In other words, the competing solvent diminishes system stability. As to the temperature of coil-helix transition, it naturally increases with the increased energy of intermolecular H-bonding.

Let us now investigate transition interval $\Delta T = -\left(\dfrac{\partial \theta}{\partial T}\right)^{-1}_{\tilde{V}+1=Q}$. Consider first the helix-coil transition. The reduced transition interval is

$$\varepsilon_{hel.-coil} = \frac{\Delta T_{comp.\,solv.}}{\Delta T_{basic\,model}} = \frac{-\left(\partial\theta/\partial\tilde{V}\right)^{-1}_{\tilde{V}+1=Q}\left(\partial\tilde{V}/\partial T\right)^{-1}_{\tilde{V}+1=Q}}{-\left(\partial\theta/\partial V\right)^{-1}_{V+1=Q}\left(\partial V/\partial T\right)^{-1}_{V+1=Q}}. \qquad (12)$$

Taking into account, that $\theta(V)$ dependence is not changed at $V \to \tilde{V}$ transformation, we obtain

$$\varepsilon_{hel.-coil,exact} = \frac{\left(\partial V/\partial T\right)_{V+1=Q}}{\left(\partial\tilde{V}/\partial T\right)_{\tilde{V}+1=Q}} = \frac{T^2_{comp.\,solv.}}{T^2_{basic\,model}} \frac{e^{E/T_{comp.\,solv.}} + q - 1}{(1-\alpha)e^{E/T_{comp.\,solv.}} + q - 1}. \qquad (13)$$

The dependence of this reduced interval on ratio of inter- and intramolecular bonding energies is presented in Fig. 3. Calculations are performed in dimensionless units $t = \frac{T}{U}$; $\alpha = \frac{mE}{U}$.

One can see, that in the $0 < \alpha < 1$ case, i.e. when the energy of intramolecular interaction is higher as compared to intermolecular one, the competing solvent diminishes the interval up to some $\alpha^* < 1$. Then the interval is growing fast. In the opposite $0 < \alpha < 1$ case, the interval increases. For coil-helix transition, in the same way, let us define

$$\varepsilon_{coil-hel.,exact} = \frac{\left(\partial\theta/\partial\tilde{V}\right)^{-1}_{\tilde{V}+1=Q}\left(\partial\tilde{V}/\partial T\right)^{-1}_{\tilde{V}+1=Q}}{-\left(\partial\theta/\partial V\right)^{-1}_{V+1=Q}\left(\partial V/\partial T\right)^{-1}_{V+1=Q}} = \frac{T^2_{comp.\,solv.}}{T^2_{basic\,model}} \frac{e^{E/T_{comp.\,solv.}} + q - 1}{(\alpha-1)e^{E/T_{basic\,model}} + q - 1}. (14)$$

One can see (Fig. 3), that increased energy of intermolecular interaction results in increased interval of coil-helix transition and this interval is smaller than that of helix-coil transition. Thus the considered model of competing solvent allows to conclude, that basing on the GMPC it is possible to investigate the influence of competing solvent on the conformational states of biopolymers. It was shown, that if the conditions $Q < q^m$ and $\frac{df_{x=0}(x)}{dx} - x_{f(x)=0} > 1$ are satisfied not only the helix-coil transition, but coil-helix transition happens as well at growing temperature. The dependencies of transition temperatures and intervals on the ratio of energies of intra- and intermolecular interactions are calculated. These results allowed to conclude, that competing solvent monotonically diminishes the temperature of helix-coil transition. The behavior of interval is more complicated: when the energy of intramolecular interaction prevails over intermolecular one competing solvent can both decrease or increase transition interval and interval – energy curve has a minimum. The coil-helix transition (ordering) happens at lower temperatures as compared to the helix-coil one and has broader interval.

The phenomenon of coil-helix transition is well-known in polypeptides and happens because at some temperatures to minimize the free energy, system breaks intermolecular H-bonds, releases solvent molecules back to solution and recovers intramolecular H-bonds.

### *3.1.2. Non-competing solvent*

Some low-weight solvent molecules reversibly bind to repeated units of polymeric chain and do not influence the formation of intramolecular H-bonds (in polypeptides such molecules bind to side chains of amino acids, in DNA they are bound to sugar phosphate backbone) [1-3]. Now let us pass to the account of such low-weight, reversibly binding non-competing solvents.

Taking into account the mechanism of influence of non-competing solvent on biopolymeric chain (non-covalent bonds with chain backbone), one can conclude, that statistical weights of conformations will change. Thus, from general ideas we suppose, that $Q$ parameter will change. Let us see, if it is right.

We define the model with non-competing solvent, basing on the following:

1. The binding of solvent molecule takes place independently from hydrogen bonding.

2. Polymer-solvent molecule bond formation depends on the orientation (state) of solvent molecule with respect to the repeated unit. These orientations are considered discrete and their number is $q$.

3. To each orientation of solvent molecule the spin variable $\alpha$, attaining values from 1 to $q$ is put into correspondence. State number one corresponds to bound state.

4. The energy of binding of non-competing solvent molecule depends on the conformation of repeated unit it is attached to. To the binding with repeated unit in state $k$ the energy of binding $E_k$ corresponds.

Taking into account all the abovementioned, we construct the Hamiltonian of model with non-competing solvent basing on the GMPC. It looks like

$$-\beta H = J\sum_{i=1}^{N}\delta_i^{(\Delta)} + \sum_{k=1}^{Q}I_k\sum_{i=1}^{N}\delta(\gamma_i,k)\delta(\alpha_i,1). \qquad (15)$$

Here $\delta(x,y)$ is Kronecker symbol; $\alpha$ is spin variable, describing the state of solvent molecule; $I_k = \dfrac{E_k}{T}$; all other designations as before.

Let us introduce the following designations

$$V = \exp[J] - 1 \Rightarrow \exp[J\delta_i^{(\Delta)}] = 1 + V\delta_i^{(\Delta)};$$

$$W = \exp[I] - 1 \Rightarrow \exp[I\delta(\gamma_i,k)\delta(\alpha_i,1)] = 1 + W\delta(\gamma_i,k)\delta(\alpha_i,1).$$

Taking into account, that $\delta(\gamma_i,\alpha)\delta(\gamma_i,\beta) = \delta(\alpha,\beta)$, the partition function is written as

$$Z = \sum_{\{\gamma_i\}}\sum_{\{\alpha_i\}}\exp[-\beta H] = \sum_{\{\gamma_i\}}\sum_{\{\alpha_i\}}\prod_i(1+V\delta_i^{(\Delta)})\prod_k\{1+W_k\delta(\gamma_i,k)\delta(\alpha_i,1)\} =$$

$$= \sum_{\{\gamma_i\}}\sum_{\{\alpha_i\}}\prod_i(1+V\delta_i^{(\Delta)})\{1+\sum_{k=1}^{q}W_k\delta(\gamma_i,k)\delta(\alpha_i,1)\} =$$

$$\sum_{\{\gamma_i\}}\prod_i(1+V\delta_i^{(\Delta)})\{1+\sum_{k=1}^{q}\frac{W_k}{q}\delta(\gamma_i,k)\} =$$

$$= \sum_{\{\gamma_i\}}\prod_i\{1+V\delta_i^{(\Delta)} + \sum_{k=1}^{q}\frac{W_k}{q}\delta(\gamma_i,k) + V\delta_i^{(\Delta)}\sum_{k=1}^{q}\frac{W_k}{q}\delta(\gamma_i,k) =$$

$$= (1+\frac{W_1}{q})\sum_{\{\gamma_i\}}\prod_i\{\frac{1+\sum_{k=1}^{q}\frac{W_k}{q}\delta(\gamma_i,k)}{1+\frac{W_1}{q}} + V\delta_i^{(\Delta)}\}$$

.

Then if make the following designations

$$x_j = \frac{1 + \sum_{k=1}^{q} \frac{W_k}{q}\delta(\gamma_j, k)}{1 + \frac{W_1}{q}}. \qquad (16)$$

The partition function is rewritten as

$$Z \propto \sum_{\{\gamma\}} \prod_i [x_i + V\delta_i^{(\Delta)}].$$

$$Z \propto \sum_{\{\gamma_i\}} [x_1 + V\delta_1^{(\Delta)}][x_2 + V\delta_2^{(\Delta)}] \cdot \ldots \cdot [x_N + V\delta_N^{(\Delta)}] =$$

$$= \sum_{\{\gamma_i\}} \{\prod_{i=1}^{N} x_i + V(\delta_1^{(\Delta)} \cdot x_1 \cdot x_2 \cdot \ldots \cdot x_N + \delta_2^{(\Delta)} \cdot x_1 \cdot x_3 \cdot x_4 \cdot \ldots \cdot x_N + \ldots) +$$
$$+ V^2(\delta_1^{(\Delta)} \cdot \delta_2^{(\Delta)} \cdot x_3 \cdot x_4 \cdot \ldots \cdot x_N + \delta_1^{(\Delta)} \cdot \delta_3^{(\Delta)} \cdot x_2 \cdot x_4 \cdot \ldots \cdot x_N + \ldots) + \ldots \}$$

As far as $x_j$, in contrast to $\delta_j^{(\Delta)}$, depends on one index, then all the terms, containing products of $x_j$, are summed independently. Thus for each sum $\sum_{i=1}^{Q} x_i = \frac{Q + \sum_{k=1}^{Q} \frac{W_k}{q}}{1 + \frac{W_h}{q}}$. In base model in place of $x_j$ were unities and the summation resulted in $\sum_{i=1}^{Q} 1 = Q$. Therefore, the difference between partition sum of model of non-competing solvent and base model consists in parameter $Q$, and the problem is reduced to base one if redefine parameter $Q$ as

$$\tilde{Q} = \frac{Q + \sum_{k=1}^{Q} \frac{\exp[I_k] - 1}{q}}{1 + \frac{\exp[I_h] - 1}{q}}. \qquad (17)$$

To analyze $\tilde{Q}(T)$ dependence we rewrite it in dimensionless units $\frac{\tilde{Q}(T)}{Q}$, $\alpha_j = \frac{E_j}{E_h}$, $t = \frac{T}{E_h}$:

$$\frac{\tilde{Q}(T)}{Q} = 1 + \frac{1}{Q} \frac{\sum_{k=2}^{Q} \exp\left[\frac{\alpha_k - 1}{t}\right] - (Q - 1)}{1 - (q - 1)e^{-1/t}}. \tag{18}$$

If simplify the problem and consider equal energies of interaction with coil, then the expression for $\tilde{Q}(T)$ is rewritten as

$$\tilde{Q} = Q + (Q - 1)\frac{\exp[\frac{E_c}{T}] - \exp[\frac{E_h}{T}]}{\exp[\frac{E_h}{T}] + q - 1}. \tag{19}$$

Schematically this dependence in $\frac{E_h}{T} = \frac{1}{t}$; $\alpha = \frac{E_c}{E_h}$ units is presented in Fig. 4.

The transition point, by analogy with base model is determined from the intersection of $\tilde{Q}(T)$ and $V(T) + 1$ curves. It is clear, that at $\alpha > 1$ the transition point is shifted to low temperatures and at $\alpha < 1$ to higher temperatures. The calculated dependence of melting temperature on $\alpha$, shown in Fig. 5 tells us the same. For the transition temperature calculations the expression

$$\tau = \frac{\beta}{1 + \ln\left[1 + (1 - 1/Q)\frac{\exp[(\alpha - 1)/\tau] - 1}{1 + (1 - q)e^{-1/\tau}}\right]}, \tag{20}$$

was used with $\beta = \frac{U}{E_h} \gg \frac{E_c}{E_h} = \alpha$; $\frac{E_h}{T}\frac{1}{\ln Q} = \frac{1}{\tau}$ designations. Thus we managed to describe the influence of non-competing solvent onto the helix-coil transition in scope of GMPC by redefinition of parameter $Q$, which becomes temperature dependent. At this if all the energies of interaction of non-competing solvent molecule with repeated unit in coil conformation are less, than the energy of interaction with helical one, then the melting temperature is higher as compared to basic GMPC (stabilization). If even one energy of interaction is higher than interaction with the helix, transition temperature is less, than in basic model (destabilization). The dependence of transition temperature on ratio between interaction energies with coil and helix is monotonically decreasing one. The qualitative picture of influence (decrease or increase) onto the helix does not depend on parameter $q$, which describes solvent entropy. Thus we conclude that non-competing solvent can either stabilize or destabilize the helical state.

Subsection 3.2. Stacking

It is widely accepted that the helical structure of DNA is conditioned by the presence of two types of interaction. The first type, known as stacking, restricts the conformational states of nearest-neighbor base pairs. It is believed that this type of interaction models the hydrophobic attraction between nearest-neighbor base pairs [1-3, 5-7]. The stacking is explained in following way: the hydrophobicity of nitrogen bases, flat heterocyclic compounds, causes parallel packing like stack of coins. The second interaction is the hydrogen bonding of complementary base pairs. Hydrogen bonding restricts the conformational states of repeated units on a finite scale, larger than nearest-neighbors'. Not only the correlation (which results in cooperativity of helix - coil transition [3]) but also the stability of DNA conformations is conditioned by these interactions. Consider how these interactions are modeled in two typical theories of the transition in polypeptides. In Zimm-Bragg theory stacking is modeled as nearest-neighbor attraction; the cost of junction between helical and coil regions [2, 23-26, 40, 41, 44] causes the cooperativity. In Lifson - Roig theory the restrictions on chain backbone conformations, imposed by hydrogen bond formation are taken into account [27-29, 33] and model the cooperativity. The results of both theories do not differ greatly [33], i.e., each factor separately results in cooperativity. To investigate the simultaneous influence of these two interactions a microscopic theory should be applied [33]. We managed to do it within the content of our Generalized Model of Polypeptide Chain (GMPC) and reveal the role of stacking against the background of hydrogen bonding.

Our base model considers cooperativity as conditioned by chain rigidity, modeled through the hydrogen bonding, while in some approaches the cooperativity is determined through so-called stacking interactions [1, 2, 5]. Taking stacking into account, by analogy with Hamiltonian Eq.(1),

$$-\beta H = J\sum_{i=1}^{N}\delta_i^{(\Delta)} + I\sum_{i=1}^{N}\delta_i^{(2)}. \tag{21}$$

The first term on the rhs is the same Hamiltonian Eq.(1), describing horizontal $\Delta$- range interactions. The second term describes nearest-neighbor-range interactions (stacking fixes in helical conformation two nearest-neighbor repeated units). Here $I=E/T$ is the reduced energy of stacking interactions. The Kronecker $\delta_i^{(2)}$ ensures that the reduced energy I is emerged when two nearest neighboring repeated units are in the same, helical conformation. The transfer - matrix for the model with the Hamiltonian Eq.(21) looks like

$$\hat{G}(\Delta) = \begin{pmatrix} VR & VR & VR & ... & VR & VR \\ R & 0 & 0 & ... & 0 & 0 \\ 0 & R & 0 & ... & 0 & 0 \\ ... & ... & ... & ... & ... & ... \\ 0 & 0 & 0 & ... & R-1 & R-1 \\ 0 & 0 & 0 & ... & 1 & Q \end{pmatrix}, \tag{22}$$

where $R=\exp[I]$; $VR=W-R$; $W=\exp[J+I]$. It is obvious, that at $R=1$ Eq.(22) passes into Eq.(2). The structure of transfer - matrix Eq. (22) is rather similar to Eq. (2). The principle difference is that in the right lower corner of $(\Delta \times \Delta)$ matrix there is some $(2 \times 2)$ matrix, which corresponds to the base model with $\Delta=2$. The secular equation for the transfer - matrix looks like:

$$\lambda^{\Delta-2}(\lambda - W)\left[\lambda^2 - (R+Q-1)\lambda + (R-1)(Q-1)\right] = R^{\Delta-1}V(Q-1). \tag{23}$$

In the same way as in the basic model, we introduce a two-particle correlation function as Eq.(4) with correlation length as Eq.(5). The calculation shows that in analogy with the basic model the temperature dependence of correlation function has a maximum. The temperature at this maximum corresponds to the transition point; and the maximal value of correlation length (Eq.

(6)) characterizes the cooperativity of transition. The transition point for the model with the Hamiltonian Eq.(21) obtained from $W=Q$ is

$$T_m = (U+E)/\ln Q.  \qquad (24)$$

Introducing $\alpha = \dfrac{E}{U}$, the energetic contribution of stacking against the background of hydrogen bonding, we calculate the dependence of the correlation length on the temperature parameter $W = \exp\left[\dfrac{U}{T}(1+\alpha)\right]$ for $\alpha$ ranging from 0 to 2. The results presented in Fig. 6 are in dimensionless units $\dfrac{\xi}{\xi_0}$, $W$; where $\xi_0$ is the maximal correlation length with $\alpha = 0$ the case, i.e. of the basic model without stacking. Fig. 7 represents the behavior of dimensionless melting temperature $\dfrac{T}{T_0}$ ($T_0$ is the melting temperature of basic model) on $\alpha$. The dependence of the dimensionless maximal correlation length on $\alpha$ is shown in Fig. 8.

In Fig. 6 one can see, that raising $\alpha$ shifts curves to the right, lowers the maxima and makes the curves wider. This shift means that the maximum of correlation length at a given $\alpha$ corresponds to values of $W>Q$. The calculations show, that the shift in the region of $\alpha \in [0,2]$ is practically linear in $\alpha$. Therefore the condition for a transition point may be written

$$\dfrac{\exp\left[\dfrac{U}{T}(1+\alpha)\right]}{Q} = 1 + c\alpha \qquad (25)$$

with constant $c>0$. In Fig. 7 one can see that melting temperature increases linearly in $\alpha$. The increase in stability is the result of stacking energy added to the Hamiltonian. The slope of this curve is less than unity. It happens due to the shifts of maximums in Fig. 6. To see the source of this slope and linearity, expand Eq.(25) around $T_0$ ($\alpha$ close to zero)

$$\dfrac{T}{T_0} = 1 + \alpha\left(1 - \dfrac{c}{\ln Q}\right). \qquad (26)$$

Fig. 7 shows, that this linearity holds up to $\alpha=2$. The linear behavior of the transition temperature explains why mean field theories [1, 2, 5, 6, 23-26], in which the helix stabilization energy is additive sum of contributions of different mechanisms of stabilization, describe the experimental data on melting temperature so well over vast range of energies.

Now consider correlations. In Fig's 6 and 8 one can see that the maximal correlation length decreases with stacking energy, i.e. $\alpha$, the increased stacking energy relative to the hydrogen bonding energy results in decreased range of correlation. At $\alpha=0$ we deal with the pure basic model with range of correlation $\Delta$; when $\alpha \to \infty$ we deal with range of correlation, equal to two, typical of stacking. As one can see from Eq.(6), the maximal correlation length for the basic model at $\alpha=0$ is much larger than at $\alpha \to \infty$. It is obvious, that at intermediate $\alpha$'s the maximal correlation length will take on intermediate values, i.e. the situation, presented in Fig. 8. So we have some mixing of maximal correlation length between the $\Delta$ - correlated and the nearest-neighbor correlated cases. This explains why increasing stacking energy vs hydrogen bonding results in the decreased correlation scale of the system.

*Section 3.3. Heterogeneity*

Natural biopolymers are systems, inhomogeneous by their structure. Proteins need to be inhomogeneous to be able to fold in special 3D structure, in which they work as enzymes. As far as between 20 aminoacids there are hydrophobic, hydrophilic and neutral ones, the correct sequence of them allows to reach this tertiary structure. As to DNA, it is known that genetic information of organism is kept in sequence of nucleotides. Thus they both are heteropolymers.

Lately we performed the account of structural inhomogeneity in heteropolypeptides, where the number of possible conformations is different for each repeated unit (each amino acid residue has different number of rotational isomers). In Refs. [53-55] it was shown, that the heterogeneity in polypeptide may be averaged with the help of redefinition of number of possible conformations of each repeated unit. In case of DNA the inhomogeneity consists in energies of hydrogen bonding and the problem can not be reduced to homogeneous case. The influence of DNA heterogeneity onto melting is known from the middle of past century and many authors investigated it theoretically and experimentally. Thus the influence of *GC* content, disorder type (random and Markovian) on $\Delta T$ и $T_m$ transition characteristics (melting temperature interval and melting point) was elucidated. However, some problems remain open. Thus, no microscopic theory of heteropolymer melting is offered. Such theory is interesting not only for biopolymer melting, but for physics of disordered systems, as far as arising problems are general [60]. The main difficulty met when constructing theory of such kind consists in partition function evaluation [2]. When using transfer-matrix formalism one faces on the problem of evaluation of trace of random product of $N$ non-commutative matrices, which is not trivial problem at large $N$. By analogy with basic GMPC let us write the Hamiltonian and transfer-matrix of heteropolymer as

$$-\beta H = \sum_{i=1}^{N} J_i \prod_{k=\Delta-1}^{0} \delta(\gamma_{i-k}, 1) = \sum_{i=1}^{N} J_i \delta_i^{(\Delta)}. \tag{27}$$

$$\hat{G}_i = \begin{pmatrix} W_i & 1 & 0 & \ldots & 0 & 0 & 0 \\ 0 & 0 & 1 & \ldots & 0 & 0 & 0 \\ \ldots & \ldots & \ldots & \ldots & \ldots & \ldots & \ldots \\ 0 & 0 & 0 & \ldots & 0 & 1 & 0 \\ 0 & 0 & 0 & \ldots & 0 & 0 & Q-1 \\ 1 & 1 & 1 & \ldots & 1 & 1 & Q-1 \end{pmatrix}, \tag{28}$$

where $\beta = T^{-1}$ is inverse temperature in energy units, $W_j = \exp[J_j] = \exp[\frac{U_j}{T}]$, where $U_j$ is the energy of hydrogen bond formation, all other symbols as in base model.

DNA in cells is synthesized taking into account the complementary feature of nitrogen bases, i.e., in the opposite of *A(T)* in one chain it is always *T(A)* in other chain and in the opposite of *G(C)* it is *C(G)*. As far as we use effective one-chain model, we no need to distinguish between chains and couples *A – T* and *T – A (G-C* and *C-G)*.

In helical state *A-T* pair is stabilized by two hydrogen bonds, while *G-C* pair by three. It means, that to break H-bonds in *j*-th repeated unit the energy $U_j$ is spent, equal to $U_{A-T}$, if in *j*-th site *A-T* pair is met, and equal to $U_{G-C}$ otherwise. $U_{A-T} < U_{G-C}$. Thus, *A-T* and *G-C* pairs differ by their energies and are undistinguishable by conformational properties. Therefore DNA is the system, inhomogeneous by the energies only and to describe it the following microscopic quantities are needed: $Q$, $\Delta$ and $\{U_i\}$ sequence. It is clear, that due to higher stability of *G-C* pairs, the higher is *G-C* content, the higher is the system stability (higher $T_m$). However, at the same *G-C* content the profile of melting curve may be different, as far as it depends on nitrogen base sequence.

The partition function at given sequence of base pairs (given disorder realisation) can be written as

$$Z_{seq} = \text{Tr} \prod_{i=1}^{n} \hat{G}_i ; \qquad (29)$$

$$\hat{G}_{AT} = \begin{pmatrix} W_{AT} & 1 & 0 & ... & 0 & 0 & 0 \\ 0 & 0 & 1 & ... & 0 & 0 & 0 \\ ... & ... & ... & ... & ... & ... & ... \\ 0 & 0 & 0 & ... & 0 & 0 & Q-1 \\ 1 & 1 & 1 & ... & 1 & 1 & Q-1 \end{pmatrix}; \quad \hat{G}_{GC} = \begin{pmatrix} W_{GC} & 1 & 0 & ... & 0 & 0 & 0 \\ 0 & 0 & 1 & ... & 0 & 0 & 0 \\ ... & ... & ... & ... & ... & ... & ... \\ 0 & 0 & 0 & ... & 0 & 0 & Q-1 \\ 1 & 1 & 1 & ... & 1 & 1 & Q-1 \end{pmatrix}.$$

As far as these matrices do not commute with each other, then every sequence from ensemble of sequences of length *n* and disorder concentration *q* has uncial statistical properties. Generally speaking, each chain can be described by sequence-dependent free energy $F_{seq}$. However, it is accepted, that free energy satisfies self-averaging principle [2, 60], which says that the distribution of probabilities of free energy values is very broad for independent sample. Therefore the free energy coincides with average free energy for almost all sequences. Of course, self-averaging takes place in thermodynamical limit, and for each particular calculation model the free energy slightly depends on sequence. However for analytical theory, investigated in thermodynamical limit, the dependence of $F_{seq}$ on sequence may be neglected at fixed disorder concentration. Thus, following Ref's [60, 61] we average the free energy by sequences. In thermodynamical limit the free energy, reduced to temeperature and number of particles is

$$f = \lim_{n\to\infty} \frac{<F_{seq}>}{nT} = -\lim_{n\to\infty} \frac{1}{n} < \ln \mathrm{Tr} \prod_{i=1}^{n} \hat{G}_i > . \qquad (30)$$

The problem of calculation of this quantity simplifies if replace the "quenched" average $<\ln \mathrm{Tr} \prod_i \hat{G}_i>$ by "annealed" one $\ln < \mathrm{Tr} \prod_i \hat{G}_i >$ at fixed disorder concentration $q$. Such replacement has basis upon it as far as thermodynamical limits of these two quantities coincide. It is the main idea of microcanonical method. Mathematical basics and background of this method are in detail discussed in book [61]. Thus it is shown, that "quenched" and "annealed" averages coincide up to the fluctuations of disorder concentration $q$, and at fixed value of $q$ these quantities are strictly equal. "Annealed" average may be expressed with the help of combinatorial methods as

$$<\mathrm{Tr} \prod_i \hat{G}_i> = \frac{1}{n_{AT}!} \binom{n}{n_{AT}}^{-1} \frac{d^{n_{AT}}}{dx^{n_{AT}}} [G_{GC} + xG_{AT}]^n \big|_{x=0} . \qquad (31)$$

The expression in square brackets may be simplified:

$$\hat{G}_{GC} + x\hat{G}_{AT} = \begin{bmatrix} W_{GC}+xW_{AT} & 1+x & 0 & 0 & \ldots & 0 & 0 \\ 0 & 0 & 1+x & 0 & \ldots & 0 & 0 \\ \ldots & \ldots & \ldots & \ldots & \ldots & \ldots & \ldots \\ 0 & 0 & 0 & 0 & \ldots & 1+x & 0 \\ 0 & 0 & 0 & 0 & \ldots & 0 & (Q-1)(1+x) \\ 1+x & 1+x & 1+x & 1+x & \ldots & 1+x & (Q-1)(1+x) \end{bmatrix} =$$

$$= (1+x) \begin{bmatrix} W(x) & 1 & 0 & 0 & \ldots & 0 & 0 \\ 0 & 0 & 1 & 0 & \ldots & 0 & 0 \\ \ldots & \ldots & \ldots & \ldots & \ldots & \ldots & \ldots \\ 0 & 0 & 0 & 0 & \ldots & 1 & 0 \\ 0 & 0 & 0 & 0 & \ldots & 0 & Q-1 \\ 1 & 1 & 1 & 1 & \ldots & 1 & Q-1 \end{bmatrix} = \hat{S}(x)\hat{\lambda}^*(x)\hat{S}^{-1}(x),$$

where $W(x)=(W_{GC}+xG_{AT})/(1+x)$; $\hat{\lambda}^*(x)=(1+x)\hat{\lambda}(x)$. Rewriting it as $[\hat{G}_{GC}+x\hat{G}_{AT}]^n=$

$$= \hat{\lambda}_1^{*n}(x)\hat{S}(x)\left[\frac{\hat{\lambda}^*(x)}{\hat{\lambda}_1^*(x)}\right]^n \hat{S}^{-1}(x),$$ and then using Frobenius-Perron theorem one can obtain

$$\left[\frac{\hat{\lambda}^*(x)}{\lambda_1^*(x)}\right]^n = \begin{bmatrix} 1 & 0 & \ldots & 0 \\ 0 & \left(\frac{\lambda_2^*(x)}{\lambda_1^*(x)}\right) & \ldots & 0 \\ \ldots & \ldots & \ldots & \ldots \\ 0 & 0 & \ldots & \left(\frac{\lambda_N^*(x)}{\lambda_1^*(x)}\right) \end{bmatrix} \xrightarrow[n\to\infty]{} \begin{bmatrix} 1 & 0 & \ldots & 0 \\ 0 & 0 & \ldots & 0 \\ \ldots & \ldots & \ldots & \ldots \\ 0 & 0 & \ldots & 0 \end{bmatrix} \equiv \hat{H}.$$

Using the last expression, we get $[\hat{G}_{GC}+x\hat{G}_{AT}]^n = \hat{\lambda}_1^{*n}(x)\hat{D}(x)$;

where $\hat{D}(x) = \hat{S}(x)\hat{H}\hat{S}^{-1}(x)$, and

$$\left\langle \mathrm{Tr}\prod_i \hat{G}_i \right\rangle = \binom{n}{n_{AT}}^{-1} \left\{ \frac{1}{n_{AT}!}\frac{d^{n_{AT}}}{dx^{n_{AT}}}[\lambda_1^{*n}(x)\hat{D}(x)] \right\}\bigg|_{x=0}. \quad (32)$$

Then using the formulae for residues

$$\text{Res } s(0) = \frac{1}{(m-1)!} \lim_{x \to 0} \frac{d^{m-1}}{dx^{m-1}} [x^m s(x)] = \frac{1}{2\pi i} \oint_R s(x) dx,$$

where $x \in C$ and $R$ is the unit circle around $x=0$ point in complex plane, one can obtain

$$<\text{Tr} \prod_i \hat{G}_i> = \binom{n}{n_{AT}}^{-1} \left\{ \frac{1}{n_{AT}!} \frac{d^{n_{AT}}}{dx^{n_{AT}}} [x^{n_{AT}+1} \left( \frac{\lambda_1^{*n}(x) \hat{D}(x)}{x^{n_{AT}+1}} \right)] \right\} =$$

$$= \binom{n}{n_{AT}}^{-1} \frac{1}{2\pi i} \oint_R \frac{\lambda_1^{*n}(x) \hat{D}(x)}{x^{n_{AT}+1}} dx = \binom{n}{n_{AT}}^{-1} \frac{1}{2\pi i} \oint_R \frac{\lambda_1^{*n}(x)}{x^{n(1-\varepsilon)}} \frac{\hat{D}(x)}{x} dx =$$

$$= \binom{n}{n_{AT}}^{-1} \frac{1}{2\pi i} \oint_R \left( \frac{\lambda_1^*(x)}{x^{(1-\varepsilon)}} \right)^n \frac{\hat{D}(x)}{x} dx = \binom{n}{n_{AT}}^{-1} \frac{1}{2\pi i} \oint_R \exp[n \ln\left( \frac{\lambda_1^*(x)}{x^{(1-\varepsilon)}} \right)] \frac{\hat{D}(x)}{x} dx.$$

From this expression with the help of steepest descent method

$$<\text{Tr} \prod_i \hat{G}_i> \approx \binom{n}{n_{AT}}^{-1} \left[ \frac{\lambda_1^*(x)}{x^{(1-q)}} \right]^n \frac{\hat{D}(x)}{x} \bigg|_{x=y}.$$

$y$ is found from extremum condition

$$\frac{d}{dy} \left[ \frac{\lambda_1^*(y)}{y^{1-\varepsilon}} \right] = \frac{d}{dy} \left[ \frac{(1+y)}{y^{1-\varepsilon}} \lambda_1(y) \right] = 0 \implies \frac{d}{d \ln y} [\ln(1+y) \lambda_1(y)] = 1-q.$$

Thus, for free energy one can obtain

$$f = -\lim_{n \to \infty} \frac{1}{n} < \ln \text{Tr} \prod_i \hat{G}_i > \approx -\lim_{n \to \infty} \frac{1}{n} \ln < \text{Tr} \prod_{i=1}^n \hat{G}_i > =$$

$$= -\lim_{n \to \infty} \frac{1}{n} \ln \left\{ \frac{1}{n_{AT}!} \binom{n}{n_{AT}}^{-1} \frac{d^{n_{AT}}}{dx^{n_{AT}}} [\hat{G}_{GC} + x\hat{G}_{AT}]^n \bigg|_{x=0} \right\} \approx$$

$$-\lim_{n \to \infty} \frac{1}{n} \ln \left\{ \binom{n}{n_{AT}}^{-1} \left( \frac{\lambda_1^*(x)}{x^{1-q}} \right)^n \frac{\hat{D}(x)}{x} \bigg|_{x=y} \right\} = = -\lim_{n \to \infty} \frac{1}{n} \ln \binom{n}{n_{AT}}^{-1} -$$

$$\lim_{n \to \infty} \ln \left( \frac{(1+x)\lambda_1(x)}{x^{1-q}} \right) \bigg|_{x=y} - \lim_{n \to \infty} \frac{1}{n} \frac{\hat{D}(x)}{x} \bigg|_{x=y}.$$

Using Stirling formulae $m! \approx m^m e^{-m}$ for the first term we obtain

$$\lim_{n\to\infty} \frac{1}{n} \ln \binom{n}{n_{AT}}^{-1} = q\ln q + (1-q)\ln(1-q).$$

The third term $\lim_{n\to\infty} \frac{1}{n} \frac{\hat{D}(x)}{x}\bigg|_{x=y}$ tends to zero due to finite $\frac{\hat{D}(x)}{x}$.

And, finally one can obtain the free energy as

$$\begin{cases} f(y) = -q\ln q - (1-q)\ln(1-q) - \ln\left[\dfrac{(1+x)\lambda_1(x)}{x^{1-q}}\right]\bigg|_{x=y} \\ \dfrac{d}{d\ln y}[\ln(1+y)\lambda_1(y)] = 1-q \end{cases} \quad (33)$$

Here $\lambda_1(y)$ is the largest eigenvalue of transfer-matrix of base model with redefined parameter $W(y) = \dfrac{W_{GC} + yW_{AT}}{1+y}$. Thus, the problem of heteropolymer is reduced to fictive homopolymeric problem.

We have an additional non-physical parameter $y$ in our fictive homopolymeric problem, which should be eliminated. We do this from the system of equations

$$\begin{cases} \dfrac{d\ln(1+y)\lambda_1(y)}{d\ln y} = 1-q \\ W(y) = \dfrac{W_{GC} + yW_{AT}}{1+y} \\ W(\lambda) = \dfrac{\lambda^\Delta(\lambda-Q)+Q-1}{\lambda^{\Delta-1}(\lambda-Q)+Q-1} \end{cases} \quad (34)$$

From here we obtain the equation of the following form, which allows to determine the largest eigenvalue of problem

$$\frac{qP_{GC} + (1-q)P_{AT}}{P_{GC} - P_{AT}} - \frac{1}{\lambda}\frac{P_{AT}P_{GC}}{(\dfrac{dP_{AT}}{d\lambda}P_{GC} - P_{AT}\dfrac{dP_{GC}}{d\lambda})} = 0, \quad (35)$$

where $P(\lambda) = \lambda^{\Delta-1}(\lambda - Q)(\lambda - W) - (Q-1)(W-1)$ is secular equation of homopolymeric (base) model. Eq. (35) allows to express $\lambda_1(y)$ through the homopolymeric parameters and calculate it. In this sense let us call this equation as heteropolymer secular equation. It can be rewritten as

$$W - \bar{W} - \frac{(W - W_{AT})(W - W_{GC})}{\lambda \frac{dW}{d\lambda}} = 0, \tag{36}$$

where $\bar{W} = qW_{GC} + (1-q)W_{AT}$. Let us investigate Eq. (36) with the help of already known information about base model. As far as $W = g(T^{-1})$ is statistical weight, then at low temperatures $W > Q$. Thus, the maximal eigenvalue of interest will be $\lambda_1 \approx W$. At high temperatures $W < Q$ and $\lambda_1 \approx Q$.

To investigate the pure influence of heterogeneity, let us consider the transition as phase. On the language of base model it means, that $\Delta \to \infty$. Let us see what happens with the heteropolymer secular equation in $\Delta \to \infty$ limit. From $W = W(\lambda)$ dependence we have

$$W = \lambda - \frac{(\lambda-1)(Q-1)}{\lambda^{\Delta-1}(\lambda-Q) + Q - 1};$$

$$\frac{dW}{d\lambda} = 1 - \frac{Q-1}{\lambda^{\Delta-1}(\lambda-Q) + Q - 1} + \frac{(\lambda-1)(Q-1)\lambda^{\Delta-2}[(\Delta-1)(\lambda-Q) + \lambda]}{[\lambda^{\Delta-1}(\lambda-Q) + Q - 1]^2}. \tag{37}$$

At low temperatures (the beginning of transition) $\frac{dW}{d\lambda} \xrightarrow{\Delta \to \infty} 1$ and from the secular equation

$$W_{lowT} = \frac{W_{AT}W_{GC}}{qW_{AT} + (1-q)W_{GC}}. \tag{38}$$

At high temperatures $\lambda_1 \approx Q$; $\frac{dW}{d\lambda} \xrightarrow{\Delta \to \infty} \infty$ and

$$W_{highT} = \bar{W} = qW_{GC} + (1-q)W_{AT}. \tag{39}$$

Let us come back to free energy and rewrite Eq. (33) as

$$f(y) = -q\ln q - (1-q)\ln(1-q) - \ln(1+y) + (1-q)\ln y - \ln\lambda_1. \qquad (40)$$

Parameter $y$ should be eliminated again. If insert two limiting values of $W$ into the expression $W(y) = \dfrac{W_{GC} + yW_{AT}}{1+y}$, then two limiting values of $y$ are obtained:

$$y_{lowT} = \frac{1-q}{q}\frac{W_{GC}}{W_{AT}}; \quad y_{highT} = \frac{1-q}{q}. \qquad (41)$$

If insert them into free energy, one can obtain

$$f(y_{lowT}) = -\ln W_{GC}^q W_{AT}^{1-q} = -\left(q\frac{U_{GC}}{T} + (1-q)\frac{U_{AT}}{T}\right); \quad f(y_{highT}) = -\ln Q. \qquad (42)$$

These low- and high-temperature approximations of free energy are in agreement with general ideas. If consider the free energy $f = \dfrac{F}{NT} = \dfrac{U}{NT} - \dfrac{S}{N} = \dfrac{u}{T} - s$ at low and high temperatures and compare with our results, then $\ln Q$ has the meaning of entropy per repeated unit and $-\ln W_{GC}^q W_{AT}^{1-q} = -\left(q\dfrac{U_{GC}}{T} + (1-q)\dfrac{U_{AT}}{T}\right)$ has the meaning of averaged internal energy.

The transition point of base model is determined from $W=Q$ equality. By analogy

$$\begin{cases} W_{highT} = q\exp\left[\dfrac{U_{GC}}{T_0}\right] + (1-q)\exp\left[\dfrac{U_{AT}}{T_0}\right] = Q, \\ W^{-1}_{lowT} = q\exp\left[-\dfrac{U_{GC}}{T_1}\right] + (1-q)\exp\left[-\dfrac{U_{AT}}{T_1}\right] = Q^{-1}, \\ W_{AT} = \exp\left[\dfrac{U_{AT}}{T_{AT}}\right] = Q, \\ W_{GC} = \exp\left[\dfrac{U_{GC}}{T_{GC}}\right] = Q. \end{cases} \qquad (43)$$

In Fig. 9 these dependencies are schematically represented. $T_0$ and $T_1$ are transition points, obtained from high- and low-temperature limits, and $T_{GC}$ and $T_{AT}$ are homopolymeric transition points. As far as the $W_{highT}$ curve is always higher, than the curve $W_{lowT}$, and they both, as averages of $W_{GC}$ and $W_{AT}$ are between them, than transition temperatures satisfy the relation $T_{AT} < T_1 < T_0 < T_{GC}$. Taking into account, that the $\left(\dfrac{1}{T_{AT}} - \dfrac{1}{T_{GC}}\right)$ quantity is small, as far as both melting temperatures are of the same order of magnitude, one can resolve the Eqs. (43) into series by small parameters $\left(\dfrac{1}{T_0} - \dfrac{1}{T_{GC}}\right), \left(\dfrac{1}{T_1} - \dfrac{1}{T_{AT}}\right)$ and obtain $T_0, T_1$. Evaluations show, that in linear approximation heteropolymer melting temperature is

$$T_0 = T_1 = T_{hetero} = qT_{GC} + (1-q)T_{AT}. \tag{44}$$

As far as $T_1$ corresponds to the beginning of transition and the $T_0$ to the end of transition, then their difference will be considered $T_0 - T_1$ as transition interval. In linear approximation these temperatures are undistinguishable, therefore to obtain the interval one should also consider the quadratic term of series. Obtained interval of heteropolymer melting looks like

$$\Delta T = T_0 - T_1 = 2q(1-q)T_{hetero} \ln Q \left(\dfrac{T_{GC} - T_{AT}}{T_{hetero}}\right)^2. \tag{45}$$

To summarize the results of this subsection we want to say, that the following results are obtained.

1) The analogue of secular equation for heteropolymer is found.

2) Basing on this equation it is shown, that the averaging regime of statistical weights changes throughout the transition.

3) The exact expression, as well as low- and high-temperature limits of free energy are obtained basing on secular equation

4) The expressions for transition temperature and interval coincide with the experimentally observable classical results. However, usually these results are postulated or just explained, but not obtained from analytical calculations, basing on exact and microscopic model as we did.

Section 4. Order parameter within GMPC.

GMPC allows to calculate the order parameter of helix-coil transition, namely, the helicity degree, which is experimentally measurable, say, through UV absorption spectrophotometry. Usually the helicity degree is defined as the average fraction of repeated units in helical conformation. Another definition of helicity degree is: the average fraction of hydrogen bonds. These two definitions are thought to be identical. However it is natural to think, that due to the important correlation of conformation in helical state it is formed not independently, but repeated units are interrelated on some scale. For example, in polypeptides three nearest neighboring repeated units should all be in helical conformation to form one hydrogen bond, which makes helical structure stable below transition temperature. Thus we think it is correct to define the helicity degree as the average fraction of hydrogen bonds. Within our GMPC we define it as the average portion of hydrogen bonds and express it through the partition function

$$Z = \sum_{\{\gamma_i\}} \exp[J \sum_{i=1}^{N} \delta_i^{(\Delta)}] \tag{46}$$

as

$$\theta = \langle \delta_i^{(\Delta)} \rangle = \frac{1}{N} \frac{\partial \ln Z}{\partial J}. \tag{47}$$

In principle, as far as we have characteristic equation Eq. (3) we can obtain eigenvalues (at least numerically) and then calculate the partition function in thermodynamic limit as

$$Z = \lambda_1^N. \tag{48}$$

It results in helicity degree as

$$\theta = \frac{\partial \ln \lambda_1}{\partial J}. \tag{49}$$

The temperature dependence of helicity degree, calculated from Eq. (49) is presented in Fig. 10. The second interesting parameter is the average portion of repeated units in helical (in our notations, number 1) conformation. This quantity is expressed through the partition function Eq. (48) as

$$\chi = \langle \delta(\gamma_i, 1) \rangle = 1 - \frac{1}{N} \frac{\partial \ln Z}{\partial \ln(Q-1)} = 1 - \frac{\partial \ln \lambda_1}{\partial \ln(Q-1)}. \tag{50}$$

The temperature dependence of this quantity, calculated from Eq. (50) is presented in Fig. 11. The comparison of curves from Fig. 10 and 11 is shown in Fig. 12.

In Fig. 10 one can see the classical behavior of helicity degree: it changes from 1 at small temperatures (completely helical state) to 0 at high temperatures (completely coil or melted state) in a very narrow temperature interval. The comparison of curves for polypeptide and DNA reveals higher cooperativity of DNA melting. Now let us observe the temperature dependence of average portion of repeated units in helical conformation. Curve a) of Fig. 11, describing the polypeptide case, shows the dependence of parameter $\chi$, identical to helicity degree dependence, while curve b) (DNA case) shows something different: parameter $\chi$ changes from 1 at low temperatures to some non-zero value at high temperatures.

We explain this difference between $\theta$ and $\chi$ with the help of following reasoning. In high temperature limit we deal with melted state and absence of correlation between conformations. It means, that all $Q$ possible conformations are equiprobable with the probability $1/Q$. From its definition $\theta$ is the probability of finding $\Delta$ nearest repeated units all in helical conformation and, thus, is equal to $Q^{-\Delta}$ in high temperature limit, while $\chi$ is the probability of finding each repeated unit in helical conformation and is equal to $Q^{-1}$ in high temperature limit. If insert polypeptide parameter set ($Q \propto 60 \div 90$; $\Delta = 3$) into these high temperature estimations of order parameters, one can see, that the probabilities $Q^{-\Delta}$ and $Q^{-1}$ are very close to each other and are about zero (see Fig. 12a). This happens due to the huge value of parameter $Q$. Thus, $\theta$ and $\chi$ have similar behavior for polypeptide case. For DNA case ($Q \propto 3 \div 5;\ \Delta \propto 10 \div 15$) the situation is contrary: the probabilities $Q^{-\Delta}$ and $Q^{-1}$ differ a lot (see Fig. 12b) from each other due to large value of parameter $\Delta$ and $\chi$ does not behave as real order parameter; it changes not between 1 and zero, but between 1 and $Q^{-1}$ instead. Basing on the above mentioned we suppose that the correct definition of the helicity degree as order parameter is: the average fraction of hydrogen bonds.

Section 5. Concluding remarks

So far we have shown that the presented approach allows to take into account factors, which affect the transition in different ways, basing on the same GMPC. The considered model describes both polypeptides and DNA (at small loop sizes), so, at presence of competing solvent the inverse (coil-helix) transition at increased temperature may happen, at least theoretically, in DNA too. This result is reported for the first time, to our knowledge, and we think, that the coil-helix transition may happen due to same reasons as in polypeptides. As far as for DNA the number of binding sites (parameter *m*) ranges from four to six and is more than *m*=1 value of polypeptide, then all the regularities are more pronounced in DNA. Thus, the interval of coil-helix transition in DNA may be within the experimental accuracy, and the transition point may be lower than 0 $C^0$, therefore not found in experiments. For the first sight it sounds strange, but it is known, that changing pressure one can rule the transition point, shifting it to positive temperatures. The transition of such kind in DNA may be experimentally found in investigations with varying pressure [62]. However, this point needs additional experimental and theoretical investigation.

Another new result is the decrease in transition cooperativity at increased stacking. This result was unexpected for us, because the term with nearest-neighbor correlation was introduced in addition to the $\Delta$ correlated one in corresponding Hamiltonian. It seemed that introducing an additional helix-stabilizing interaction would result in an increased maximal correlation length as well as it resulted in increased melting temperature, but it was not so. It is widely accepted that the main cause of cooperativity is stacking. However it should be kept in mind, that the comparative analysis of two cooperativity factors, stacking and hydrogen bonding, can not be performed with the frequently used mean field approximation. So we must compare our results with those of another microscopic theory of DNA melting which takes into account the difference in the mechanisms of hydrogen bonding and stacking. Consider another microscopical approach, namely, the Peyrard-Bishop approach [11-13, 51]. This model recognizes both stacking and hydrogen bonding. It uses the analogy between the statistical treatment of macromolecules and the Schroedinger equation [2, 63] and treats the problem as a particle in Morse potential which reflects hydrogen bonding. Harmonic coupling reflecting stacking is assumed between nearest neighbor repeated units. The case of disorder in sequence [32] was studied as well. The increase in melting temperature and the decrease in melting interval follows from Fig.1 of Ref.[13]. This Figure compares the calculated melting curves for different contributions of stacking. The authors discuss only melting temperature and compare it with Zimm-Bragg mean-field theory. In the same picture the increased melting interval at increased stacking is shown, but there are no comments concerning this point.

The Peyrard-Bishop results coincide with ours. In both models the difference of mechanisms of stacking and hydrogen bonding interactions in DNA is correctly introduced. In the Peyrard-Bishop approach this difference is included as the difference of interaction potentials. For example to reach the better qualitative accordance with experiment they introduce the anharmonicity of interaction potential. In our model this difference consists in the different correlation scales, as directly follows from DNA structure. As we showed above the action of water and other solvents may be included in our basic model by redefinition of model parameter *W* or *Q*. The choice depends on the mechanism of solvent interaction. It means that our qualitative analysis is applicable to DNA melting experiments. As is known from the literature, the contributions of stacking and hydrogen bonding into the energy of helical state are of the same order in water [59]. Therefore in real DNA we deal with the case $\alpha \sim 1$, which corresponds to significant lowering of the maximal correlation length compared to the case $\alpha = 0$ (pure basic model). This lowering is by twenty times (Fig. 8). So we suggest that the increased role of stacking (or the decreased role of hydrogen bonding) will result in decreased cooperativity. It may be interesting to check these results experimentally. For example one might add flat cyclic compounds such as porhyrins that can stack in DNA major groove thus increasing the stacking in system, or to add some hydrophilic solvent, then melt DNA and investigate the change in temperature interval.

As to the heteropolymer problem, the approach used (microcanonical trick) belongs to new mathematical methods, developed in last decade. It was believed, that no analytical estimations could be performed using this method, only numerical ones instead [61]. We succeed to estimate quantities of interest analytically and the results obtained for melting temperature and interval coincide with classical ones. Moreover, till now many heteropolymer theories were based on the ansatz (see Ref. 1) about the disorder distribution and averaging type; whether it is arithmetic or harmonic mean of two possible coupling energies. We found without any ansatz about the averaging type, that the averaging of statistical weights corresponding to two types of coupling changes throughout the transition: from harmonic mean at low temperatures to arithmetic one at high temperatures. The results obtained are important not only for statistical physics of macromolecules, but for physics of disordered systems in whole [60]. In particular we managed to map the considered problem to the problem of Anderson localization of electron and are now working in this direction as well.

And finally we want to discuss the comparison of theoretical results with experimental data. At first we accept, that no quantitative coincidence is reached, only qualitative instead. It happens due to several reasons. Thus, for example, the exact, calculated values of energies of interactions are unknown. Of course, they have been measured experimentally and therefore contain mixed information, not pure one. For example, the measured energy of hydrogen bonding contains the energy of interaction with water and many, many others. Moreover, we think, that no quantitative coincidence should be expected, as far as we do not introduce any fitting parameters in our theory. Instead, being based on non-mean-field theory, the approach allows to investigate the mechanism of influence of these factors on helix-coil transition. Thus, it could not be revealed, that stacking decreases the transition cooperativity without such a microscopic theory. Of coarse many open questions exist. For example, as far as biopolymers both *in vivo* and *in vitro* are functioning in presence of different high molecular weight compounds, the influence of them onto conformational transitions should be taken as well. This point and many others, like DNA-DNA interaction is under consideration now, and we hope that these results will be published soon.

Acknowledgments

The partial support of ICTP is thankfully acknowledged by three of authors: A.V. Badasyan, E.Sh. Mamasakhlisov, V.F. Morozov. The work of E.Sh.M. and V.F.M. was also partly supported by ISTC grant No. A-301.2. The ANSEF Grant 05-ns-molbio-815-503 is also thankfully acknowledged.

References

1. Poland, D.C.; Scheraga, H.A. The Theory of Helix-Coil Transition; Academic Press: New York, 1970.

2. Grosberg, A.Yu.; Khokhlov, A.R. Statistical Physics of Macromolecules; AIP Press: New York, 1994; Chapter 7.

3. Cantor, C.R.; Shimmel, T.R. Biophysical Chemistry; Freeman and Co.: San-Francisco, 1980; Chapter 20.

4. Flory, P.J. Statistical Mechanics of Chain Molecules; Interscience: New York, 1969; Chapter 7.

5. Wartell, R.M.; Benight, A.S. Phys. Rep. 1985, 126 (2), 67-107.

6. Vedenov, A.A.; Dykhne, A.M.; Frank-Kamenetskii, M.D. Usp Phys Nauk (in Russian) 1971, 105, 479-519.

7. Wada, A.; Suyama, A. Prog Biophys Mol Biol 1986, 47, 113-157.

8. Fixman, M.; Zeroka, D. J. Chem. Phys. 1968, 48, 5223.

9. Volkenstein, M.V. in Configurational Statistics of Polymeric Chains: Wiley Interscience 1963.

10. Dashevskii, V. G. in Conformational Analysis of Macromolecules: Moscow Publ. 1987.

11. Peyrard, M.; Bishop, A. R. Phys. Rev. Lett. 1989, 62(23), 2755-2758.

12. Dauxois, T.; Peyrard, M.; Bishop, A.R. Rap Comm Phys Rev E 1993, 47(1), 44-47.

13. Dauxois, T.; Peyrard, M.; Bishop, A.R. Phys Rev E 1993, 47(1), 684-695.

14. Frank-Kamenetskii, M. D. Mol. Biol., 1968, 2, 408.

15. Frank-Kamenetskii, M. D.; Karapetyan, A. T. Mol. Biol. 1972, 6, 621.

16. Melchior, W.B.; von Hippel, P.H. Proc. Nat. Acad. Sci. U.S., 1973, 70, 298.

17. Go, N.; Go, M.; Scheraga, H.A. J. Chem. Phys. 1971, 54, 4489.

18. Go, N.; Go, M.; Scheraga, H.A. Macromolecules 1974, 7, 459.
33

Figure 1.

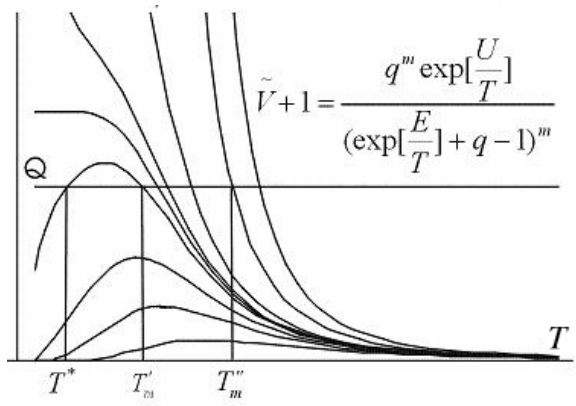

Figure 2.

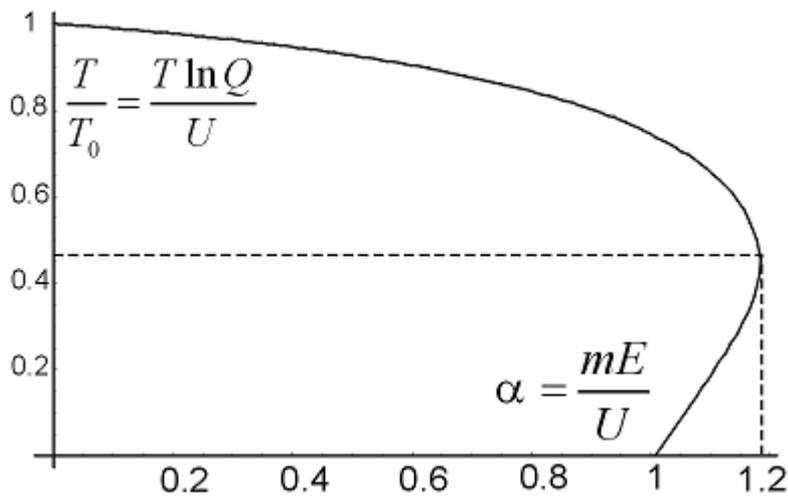

Figure 3.

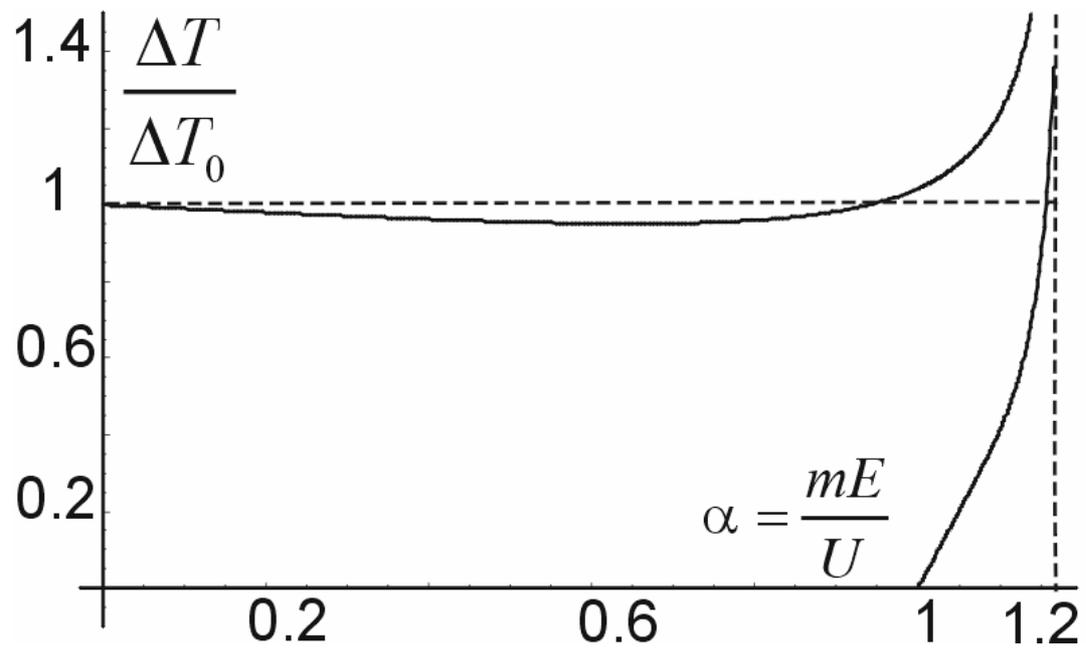

Figure 4.

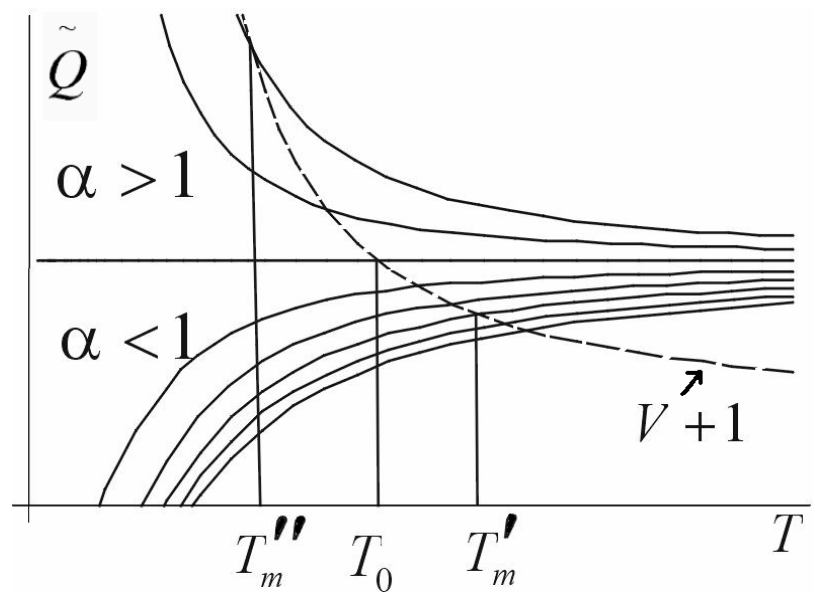

Figure 5.

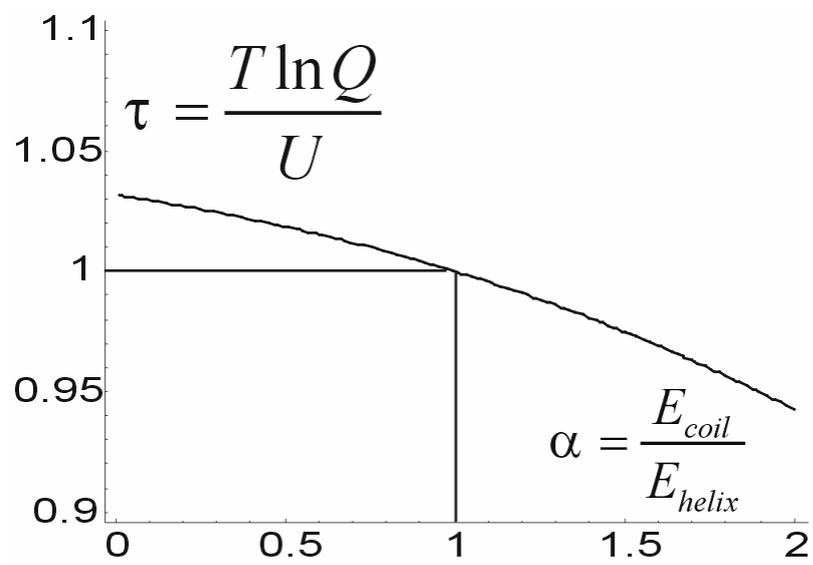

Figure 6.

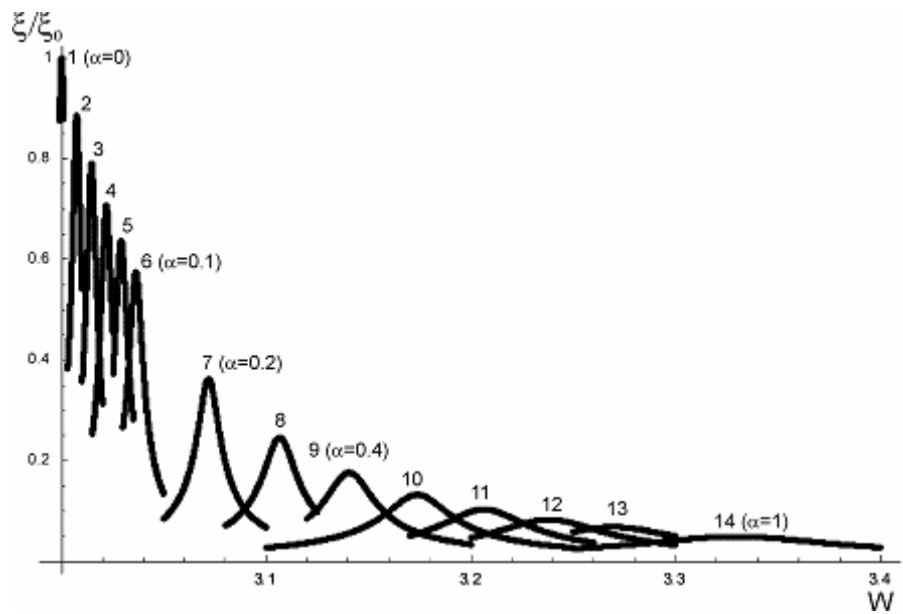

Figure 7.

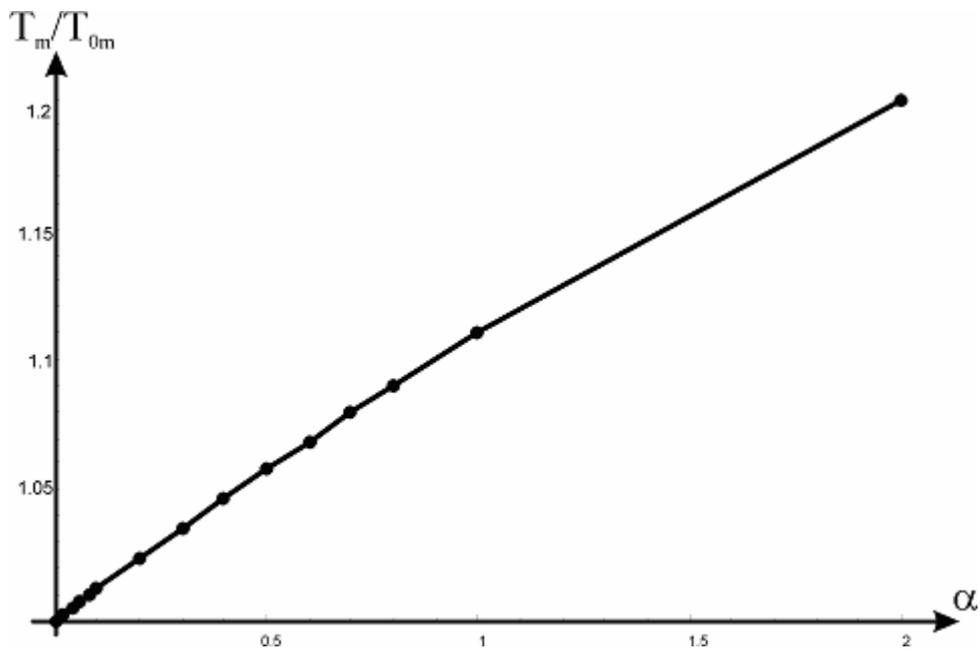

Figure 8.

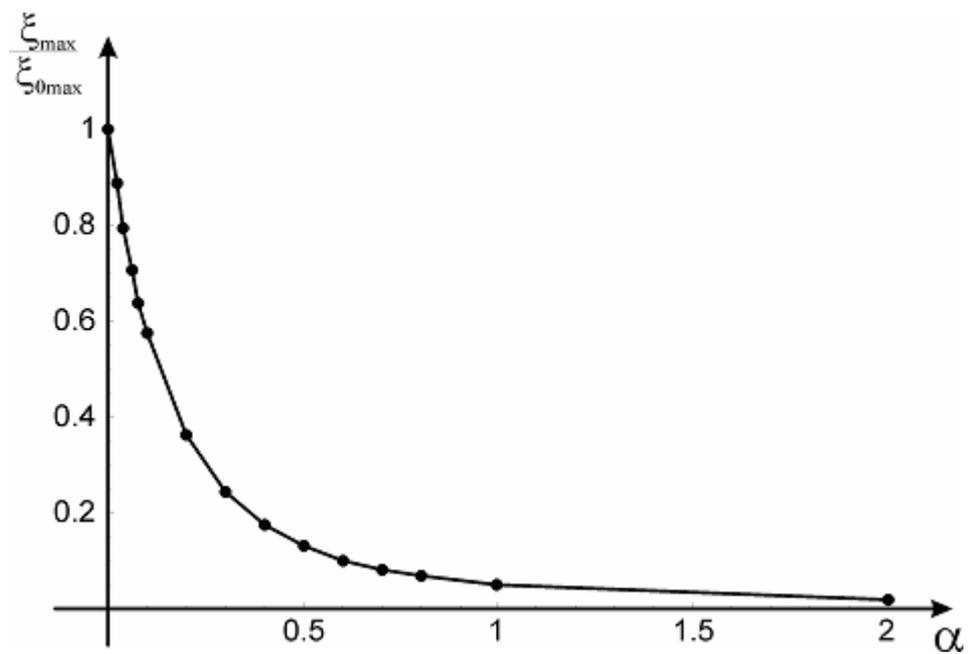

Figure 9.

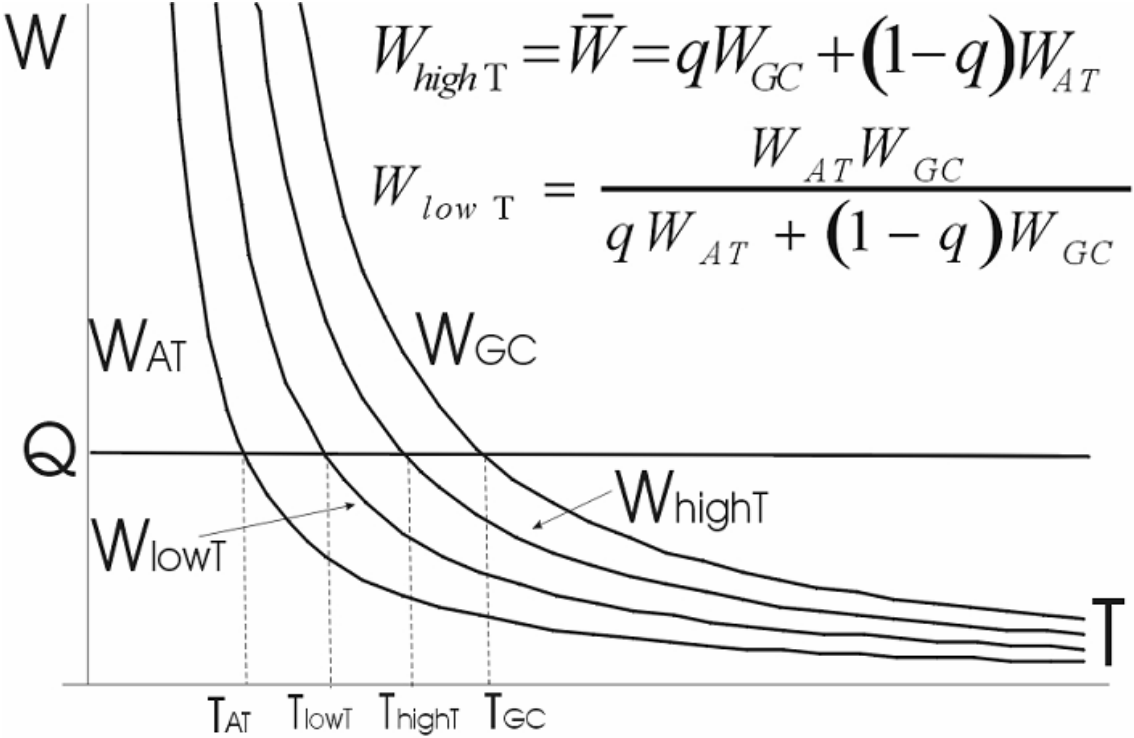

Figure 10.

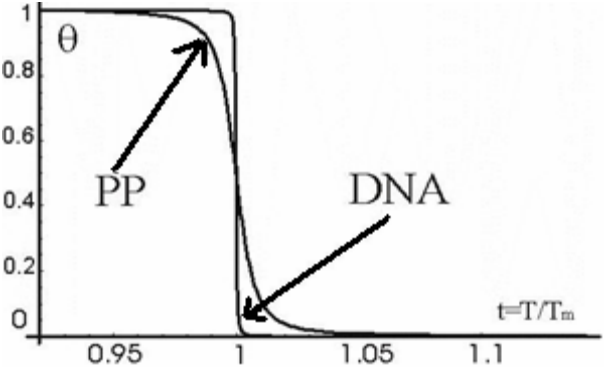

Figure 11.

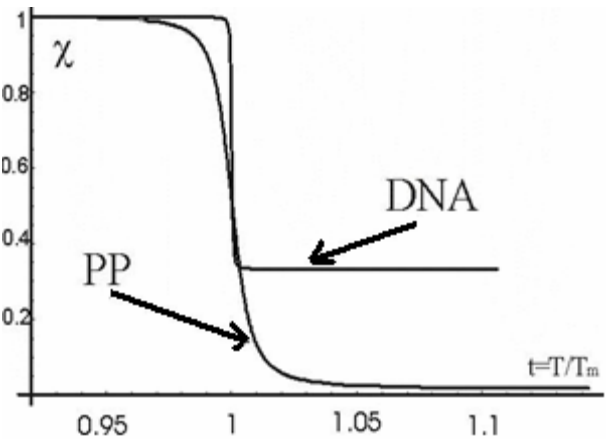

Figure 12 a.

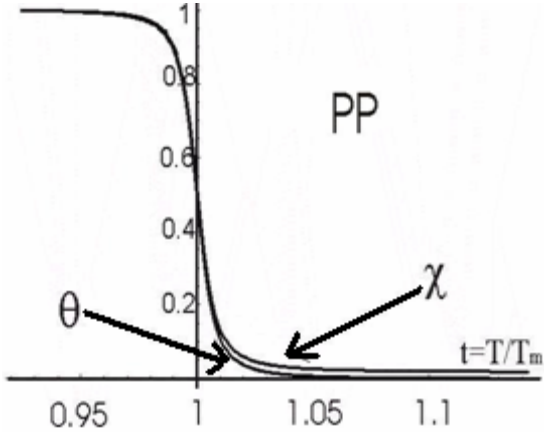

Figure 12 b.

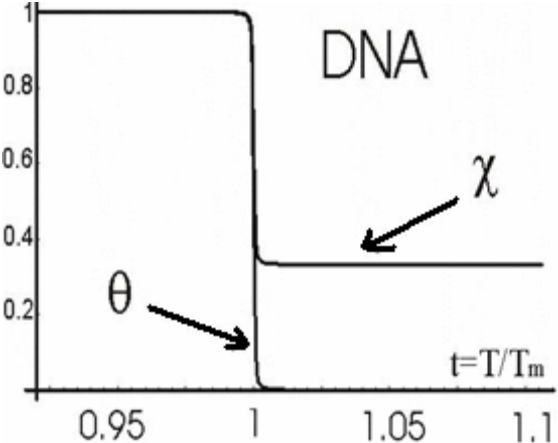

Figure Captions

Figure 1. The temperature dependences of $\tilde{V}(T)+1$ at different values of $\alpha = \dfrac{mE}{U}$ parameter.

Figure 2. The dependence of reduced transition temperature on energy parameter for helix-coil and coil-helix transitions in dimensionless units at presence of competing solvent.

Figure 3. The dependence of reduced transition interval on energy parameter for helix-coil and coil-helix transitions in dimensionless units at presence of competing solvent.

Figure 4. The temperature dependence of redefined parameter $\tilde{Q}(T)$.

Figure 5. The dependence of reduced transition temperature on energy parameter for helix-coil transition in dimensionless units at presence of non-competing solvent.

Figure 6. The dependencies of reduced correlation lengths $\xi/\xi_0$ of model with stacking on temperature parameter $W$ for fixed values of $\alpha = \dfrac{E}{U}$. The curves are enumerated corresponding to the following values of $\alpha$: 1 – 0; 2 – .002; 3 – .004; 4 – .006; 5 – .008; 6 – .1; 7 – .2; 8 – .3; 9 – .4; 10 – .5; 11 – .6; 12 – .7; 13 – .8; 14 – 1.

Figure 7. The dependence of reduced maximal correlation length $\xi_{max}/\xi_{0\,max}$ on $\alpha$.

Figure 8. The dependence of reduced transition temperature $T_m/T_{0m}$ on $\alpha$.

Figure 9. The scheme, explaining temperature dependence of different averagings of temperature parameter $W$ in heteropolymer model.

Figure 10. The temperature dependence of $\theta$ helicity degree for polypeptide (PP) and DNA. Temperature is reduced to melting temperature.

Figure 11. Temperature dependence of $\chi$ parameter for polypeptide (PP) and DNA. Temperature is reduced to melting temperature.

Figure 12 a. The temperature dependence of both $\theta$ and $\chi$ for polypeptide parameter set.

Figure 12 b. The temperature dependence of both $\theta$ and $\chi$ for DNA parameter set.